\documentclass[11pt,fleqn]{article}
\usepackage[top=0.8in,bottom=0.7in,left=0.2in,right=0.2in,headheight=14pt,headsep=10pt,columnsep=14pt]{geometry}

\usepackage{amsmath,amssymb}
\usepackage{bm}
\usepackage{graphicx}
\usepackage{titlesec}
\usepackage{algorithm}
\usepackage{algorithmic}
\usepackage{listings}
\usepackage[svgnames]{xcolor}
\usepackage{caption}
\usepackage{etoolbox}

\AtBeginEnvironment{equation}{\noindent}
\AtBeginEnvironment{equation*}{\noindent}
\usepackage{fancyhdr}
\usepackage[numbers,square]{natbib}

\usepackage{listings}
\usepackage[svgnames]{xcolor}
\usepackage{hyperref}
\hypersetup{
    colorlinks=true,
    linkcolor=blue,
    citecolor=blue,
    urlcolor=blue,
    filecolor=blue
}
\usepackage{cleveref}
\usepackage{caption}
\usepackage{footmisc}
\usepackage{authblk}
\usepackage{multicol}

\titlespacing*{\section}{0pt}{20pt}{10pt}
\titlespacing*{\subsection}{0pt}{16pt}{6pt}
\titlespacing*{\subsubsection}{0pt}{8pt}{3pt}

\titleformat*{\section}{\large\bfseries}
\titleformat*{\subsection}{\normalsize\bfseries}
\titleformat*{\subsubsection}{\normalsize\bfseries}

\fancypagestyle{plain}{%
    \fancyhf{}%
    \fancyhead[L]{\hspace*{0.3in}R. Meng and X. Dong (2026)}%
    \fancyhead[R]{\thepage\hspace*{0.3in}}%
}

\newcommand{\grad}{\nabla}

\newcommand{\etal}{et~al.}

\DeclareMathSizes{10}{9.5}{7}{5}

\newenvironment{keywords}{%
  \begin{quote}\noindent\textbf{Keywords:}~%
}{\end{quote}}

\begin{document}

\title{Wave Cold Dark Matter Can Reproduce the Radial Acceleration Relation in X-ray Elliptical Galaxies}

\author[a,b]{Run-Yu Meng}
\author[a,b]{Xiao-Bo Dong\thanks{Corresponding authors. Email: xbdong@ynao.ac.cn\\}}

\affil[a]{Yunnan Observatories, Chinese Academy of Sciences, Kunming, Yunnan 650216, China}
\affil[b]{University of Chinese Academy of Sciences, Beijing 100049, China}

\date{\today}

\maketitle

\abstract{
    \setlength{\baselineskip}{16pt}%
	Wave cold dark matter (wave CDM) is a well-motivated variant within the collisionless CDM paradigm, 
	in which the de~Broglie wavelengths of ultralight bosons become relevant on galactic scales.
	While many of the previous tests of wave CDM have focused on
	small-scale tensions involving solitonic cores and certain soliton--halo relations in galaxies, 
	the radial acceleration relation (RAR) provides a complementary test: 
	for a given baryonic mass distribution, it probes whether the DM halo can reproduce the observed total gravitational field.  
	In this work we test wave CDM against the RAR using eight X-ray elliptical galaxies
	from the compilation of \mbox{Lelli \textit{et al}}.  
    We perform a suite of three-dimensional Schr\"odinger--Poisson simulations
    and evolve the wave-CDM halos to virialized quasi-stationary states.
    For each galaxy and each adopted boson mass, 
    we select the realization whose final snapshot most closely matches the empirical RAR.
    We find that the selected wave-CDM realizations for these elliptical galaxies
    can reproduce the RAR without including baryonic feedback or dark-matter self-interactions.	
    %%%
	At the boson mass $m_a=10^{-23}\,\mathrm{eV}$, the combined fit to the eight galaxies gives
	$a_0=1.19\times10^{-10}\,\mathrm{m\,s^{-2}}$, 
	consistent with the canonical value inferred from disk galaxies.  
    When the boson mass is varied over $3\times 10^{-24}\,\mathrm{eV} \le m_a \le 2\times 10^{-23}\,\mathrm{eV}$, 
    the best-fit value of $a_0$ changes by only $\sim 11\%$, 
    and this variation shows no systematic trend with $m_a$.
	%%%
	This insensitivity to boson mass arises
	because, over the radial range relevant for the RAR, 
	the virialized halos are supported mainly by their velocity field, 
	whereas the boson-mass-dependent quantum-pressure term is dynamically subdominant.
	Consequently, different boson masses can yield similar acceleration relations 
	despite producing different central solitonic structures.  
	These results show that wave CDM can reproduce the RAR in elliptical galaxies, 
	and provide a new galactic-scale test of the model beyond the previous focus on solitonic cores.
}


\begin{multicols}{2}

\section{Introduction} \label{sec:intro}

Wave cold dark matter (also known as fuzzy dark matter)%
\footnote{Related models appear in the literature under a variety of partly overlapping names, 
	including scalar-field, ultralight bosonic, axion-like (ALP), and Bose--Einstein-condensate dark matter (BEC DM).  
	These descriptions reflect the treatment of DM as a highly occupied ultralight scalar field, 
	which may exhibit coherent wave-like, condensate-like, or superfluid-like behavior under appropriate conditions.  
	The term ``superfluid DM'' has more than one origin in the literature.  
	Besides its occasional use in connection with fuzzy DM, 
	it also commonly denotes a distinct, MOND-motivated class of models featuring
    non-negligible DM self-interactions and couplings between DM and baryons. 
    Strictly speaking, it should therefore not be used synonymously with the collisionless wave-CDM model considered here.
  } %%% July14.
is a variant within the collisionless cold dark matter (CDM) paradigm.
Originally proposed as a way to address the small-scale problems of conventional particle CDM
\citep{hu2000fuzzy},
it models DM as ultralight bosons with masses typically in the range
$m_a \sim 10^{-24}$--$10^{-20}\,\mathrm{eV}$.
For such small particle masses, the de~Broglie wavelength,
$\lambda_{\mathrm{dB}} \equiv {h}/{(m_a v)}$,
can become astrophysically relevant inside galactic halos.
The resulting nonlinear wave dynamics can then alter the internal structure of DM halos 
(see \cite{hui2021wavedark} and the references therein),
thereby providing a route toward resolving long-standing small-scale challenges in $\Lambda$CDM
(see \cite{bullock2017smallscale} for a review).

Among the challenges, 
a particularly puzzling aspect of galaxy phenomenology is 
the appearance of tight scaling relations between dynamical and baryonic properties. 
A well-known example is the baryonic Tully--Fisher relation (BTFR; \cite{mcgaugh2012baryonictully}).
More recently, \cite{mcgaugh2016radial} introduced a generalized form of BTFR, the radial acceleration relation (RAR),
which correlates the observed total gravitational acceleration $g_\mathrm{obs}$ 
with the baryonic Newtonian acceleration $g_\mathrm{b}$ inferred from ordinary matter at each radius:
\begin{equation} \label{eq:RAR}
    g_{\mathrm{obs}} = \frac{g_\mathrm{b}}{1 - e^{-\sqrt{g_\mathrm{b}/a_0}}} \, .
\end{equation}
Here, $a_0 \approx 1.20 \times 10^{-10} \, \mathrm{m\,s^{-2}}$ is the characteristic acceleration scale.
For $g_{\mathrm{b}} \ll a_0$, expanding Eq.~\ref{eq:RAR} gives 
$g_{\mathrm{obs}} \sim \sqrt{g_{\mathrm{b}}a_0}$; 
for $g_{\mathrm{b}} \gg a_0$, the exponential term is negligible, giving $g_{\mathrm{obs}} \sim g_{\mathrm{b}}$.
Equation~\ref{eq:RAR} thus interpolates smoothly and monotonically between these two limits.
The RAR is, in fact, a tight correlation between 
the DM contribution $g_{\mathrm{DM}} \equiv g_{\mathrm{obs}} - g_{\mathrm{b}}$ 
and the baryonic contribution $g_{\mathrm{b}}$ at the same location.
The tightness of this empirical relation suggests that, at least phenomenologically,
baryons and the dark sector are coupled more strongly than expected from purely gravitational dynamics
within standard $\Lambda$CDM \citep{bullock2017smallscale}.

From the standpoint of wave CDM, such tight baryon--DM correlations should arise from 
bosonic physics (including Bose--Einstein condensation and superfluid effects),
the associated wave dynamics and formation processes. %July9.
%% rather than from an additional, explicitly engineered interaction. 
Unfortunately, only a few studies have pursued this direction. 
In the existing literature, reproducing the observed scaling relations 
typically requires adding ingredients beyond the collisionless CDM picture, 
often in the form of a finely tuned non-gravitational interaction mechanism \citep[e.g.,][]{hossenfelder2020milky}.
Therefore, the question of whether the observed baryon--DM tight correlations 
can emerge without introducing extra mechanisms remains open.

On the other hand, a large body of work has used rotation curves 
to constrain the boson mass and, when relevant, the strength of boson self-interactions in ultralight DM models
\citep[see, e.g.,][and the references therein]{bar2022galactic,indjin2026dynamicalreconstruction}.
It was found that a single $m_a$ value cannot account for the rotation curves across different galaxies.
Interpreted in a broader modeling context, this suggests that additional physical effects may be required,
such as self-interactions or a certain tailored finite-temperature contribution \citep[e.g.,][]{bernal2018rotation}.

But there is a key complication in those rotation-curve studies:
the analyses rely on certain assumed \emph{universal} soliton--halo relations, such as 
a tight mass relation between the solitonic core and the surrounding halo,
the scaling relations in the soliton and the total halo profiles
\citep{robles2019scalar}.
%%%%
More recently, however, there is growing evidence that the soliton--halo connection need not be universal.
Self-consistent, stable halos can be constructed to have a variety of density profiles,
with different core--halo mass relations (e.g., \cite{yavetz2022construction}).
In fact, the mass relations vary with the dynamical state of the halo 
(i.e., depending on the internal motions within the halo, namely the \emph{velocity field}; 
see \cite{liao2025decipheringsolitonhalo}).

%% our work:
Motivated by these considerations, the present work asks 
whether wave CDM --- without dark-sector self-interactions or any non-gravitational baryon--DM coupling --- can reproduce 
the observed tight baryon--DM correlations on galactic scales. 
Here we focus on the RAR (Eq.~\ref{eq:RAR}), 
which does not require partitioning the DM density distribution %$\rho_\mathrm{DM}(r)$ 
into a central solitonic core and an outer halo, 
and thus does not assume any specific soliton--halo mass relation. 	
It therefore provides a complementary and \emph{independent} way to test wave CDM.
%%%%%%
\cite{lelli2017one} showed that this same relation describes a broad variety of galaxies,
including both rotation-supported disk galaxies and pressure-supported ellipticals.
As a first step, in this work we consider the eight X-ray luminous elliptical galaxies in their sample. % of \cite{lelli2017one}.
These systems are massive, gas-poor, structurally simple, and approximately spherical,
making them a relatively clean test bed.

%% new para, Runyu July8:
In this study, we perform numerical simulations of these eight elliptical galaxies with two main goals.
First, we examine whether virialized wave-CDM halos can reproduce the empirical RAR
without invoking any additional physics such as baryonic feedback due to outflows, 
dark-matter self-interactions, etc.
Second, we test whether the resulting relation, and in particular the best-fit value of $a_0$, 
depends on the boson mass, which we vary over one order of magnitude for each galaxy.

The remainder of this paper is organized as follows. 
Sect.~\ref{sec:theory} introduces the theoretical framework of wave-CDM halos. 
Sect.~\ref{sec:numerical} describes the numerical methods used to initiate and evolve the galactic halos. 
Sect.~\ref{sec:results} presents the simulation results for the eight elliptical galaxies, 
with particular emphasis on the RAR and its insensitivity to the boson mass. 
Sect.~\ref{sec:conclude} summarizes our main findings and discusses their implications. %%July9night,xbdong.

\section{Theoretical Background}\label{sec:theory}

Being composed of ultralight bosons,
the most fundamental characteristic of wave CDM in galactic-scale halos
is the extremely high critical temperature for Bose--Einstein condensation,
$T_c \sim 10^{35}\,\mathrm{K}$ \citep{chavanis2019predictive}.
By comparison, the cosmic microwave background temperature is approximately $3\,\mathrm{K}$\,, 
and the effective temperature associated with the kinetic energy of the bosons is only
$T_{\mathrm{eff}} \sim 10^{-25}\,\mathrm{K}$ for the fiducial boson mass $m_a \sim 10^{-22}\,\mathrm{eV}$\,.
%%\citep{chavanis2019predictive}.
Thus, relative to $T_c$\,, the system is effectively in the zero-temperature regime.

%%% July 12:
In this regime, the dynamics of a highly Bose-degenerate bosonic system 
can be described by a complex-valued macroscopic wave function
$\psi(\mathbf r,t)$, which plays the role of the order parameter. 
Its evolution is governed by the Gross--Pitaevskii equation (GPE); 
see \cite{pitaevskii2016boseeinstein} for the theoretical background.
%% Once the system reaches a ``quasi-equilibrium'' state, 
For a virialized self-gravitating halo, however, 
this wave function is not expected to represent a pure ground state. 
Instead, the system generally consists of a coherent central, ground-state-like component 
embedded in an extended population of excited modes.  
The former is commonly referred to as the ``soliton'' in the wave-CDM literature, 
while the latter carries collective excitations, granular structures, 
and some turbulent non-condensed components
(see, e.g., \cite{liu2023coherent} for numerical analyses, 
and \cite{chiueh2023theoryphenomenology, proukakis2023unifieddescription} for theoretical studies).

%% new para, July 12, xbdong added:
The use of a macroscopic wave function in this context 
does not require the entire system to have established global phase coherence.  
In the highly Bose-degenerate, effectively zero-temperature regime considered here, 
the de~Broglie wavelengths of the bosons strongly overlap, and the relevant low-energy modes are highly occupied.  
The field $\psi(\mathbf r,t)$ can therefore be interpreted as a classical matter-wave field 
describing these highly occupied bosonic modes; 
see \cite{kagan1997evolutioncorrelation} for a detailed discussion.

\subsection{The Schr\"odinger--Poisson equation}

For a self-gravitating halo,
the GPE and Poisson equation form a coupled system
called the Schr\"odinger--Poisson equation (SPE):
\begin{eqnarray} 
\label{eqn:GPE}
    i\hbar \frac{\partial\psi}{\partial t}(\mathbf{r},t) = -\frac{\hbar^2}{2m}\grad^2 \psi(\mathbf{r},t) + m\Phi_{\mathrm{total}}(\mathbf{r},t)\psi(\mathbf{r},t) \,, \\
%% \end{eqnarray}
%% \begin{equation} 
\label{eqn:Poisson}
    \grad^2 \Phi_{\mathrm{total}}(\mathbf{r},t) = 4\pi G\left[|\psi(\mathbf{r},t)|^2 + \rho_{\mathrm{baryon}}(\mathbf{r},t)\right]\,,
\end{eqnarray}
where $\psi(\mathbf{r},t)$ denotes the macroscopic wave function,
$m$ represents the boson mass,
and $\Phi_{\mathrm{total}}$ is the total gravitational potential from both the wave CDM density $|\psi|^2$ and the baryonic density $\rho_{\mathrm{baryon}}$\,.

Applying the Madelung transformation provides a hydrodynamic representation of the Schr\"odinger--Poisson system, as follows.
Expressing the wave function in polar form,
\begin{equation} \label{eqn:modulus-arg}
    \psi(\mathbf{r},t) = \sqrt{\frac{\rho(\mathbf{r},t)}{m}} \, e^{iS(\mathbf{r},t)/\hbar}\,,
\end{equation}
where $\rho(\mathbf{r},t) = m|\psi(\mathbf{r},t)|^2$ represents the density field
and $S(\mathbf{r},t)$ is the phase function, one obtains the velocity field
\begin{equation} \label{eqn:fluid_velocity_field}
    \mathbf{u}(\mathbf{r},t) = \frac{1}{m}\grad S \,.
\end{equation}
%%%
Substituting Eq.~\ref{eqn:modulus-arg} into the Schr\"odinger equation (Eq.~\ref{eqn:GPE}) and separating real and imaginary parts, we obtain
\begin{equation} \label{eqn:Mad_continuity}
    \frac{\partial\rho}{\partial t}+\grad\cdot (\rho \mathbf{u})=0 \,,
\end{equation}
\begin{equation} \label{eqn:Mad_Euler}
\frac{\partial \mathbf{u}}{\partial t} +(\mathbf{u}\cdot \grad)\mathbf{u} =  -\grad\Phi_{\mathrm{total}} -\frac{1}{m} \grad Q \,,
\end{equation}
where $Q$ denotes the quantum potential (also called Bohm potential):
\begin{equation} \label{eqn:Potential_Q}
Q=-\frac{\hbar^2}{2m}\frac{\grad^2 \sqrt{\rho}}{\sqrt{\rho}}=-\frac{\hbar^2}{4m}\left[ \frac{\grad^2\rho}{\rho}-\frac{1}{2}\frac{(\grad\rho)^2}{\rho^2}\right] \,,
\end{equation}
its gradient is called quantum pressure, 
and Equation~\ref{eqn:Mad_Euler} is usually called the quantum Euler equation.

\subsection{Virialization and quasi-stationary states} \label{sec:virialization}

It is well known that three-dimensional self-gravitating systems cannot attain thermodynamic equilibrium
\citep{lynden-bell1967statisticalmechanics}.
After mean-field relaxation,
such systems become trapped in non-equilibrium quasi-stationary states
(NEQSSs, \cite{benetti2014nonequilibrium}).
These NEQSSs, which are overall virialized, sustain weak (wave) turbulence inherited from the mean-field dynamics
such as the ``violent relaxation'' during gravitational collapse
\citep{klessen2000gravitational,berman2012turbulence,moon2024theory}.
By and large, they can be approximately regarded as \emph{statistically stationary} turbulence,
and thus their global velocity field has been routinely represented as an effective velocity-dispersion pressure term
\citep{chandrasekhar1953problems,klessen2000gravitational}.
In the case of wave-CDM halos in NEQSSs, recently we provided a complete Jeans-like equation 
involving gravitational force, quantum pressure, 
and the radial profiles of angle-averaged mean-velocity and velocity-dispersion terms \citep{meng2026wavedmjladaptablea}.

We assess whether each halo has reached a NEQSS using three diagnostics 
as implemented in the \texttt{WaveDM.jl} package \citep{meng2026wavedmjladaptablea}:
(1) evolution of overall energy terms,
(2) stationarity of the radial mass distribution in terms of mass-fraction radii, and
(3) the above-mentioned Jeans-like equation indicative of NEQSSs.
A detailed description of these diagnostics is provided in Sect.~\ref{sec:app_virial}.

\section{Numerical Methodology}\label{sec:numerical}

Simulations of wave CDM impose stringent resolution requirements:
accurately modeling wave phenomena on galactic scales requires numerical tools
capable of simultaneously capturing the global structure of DM halos
and resolving the associated wave patterns on spatial scales much smaller than
the local de~Broglie wavelength
\citep[see, e.g.,][]{li2019numerical,schive2026fuzzydark}.
A variety of numerical methods and software frameworks have been developed
for simulating the Schr\"odinger--Poisson system in the context of
wave CDM, %ultralight bosonic dark matter,
including adaptive-mesh-refinement (AMR) methods
\citep[e.g.,][]{schive2014cosmicstructure},
pseudo-spectral split-step Fourier methods (SSFM)
\citep[e.g.,][]{schwabe2016simulations, mocz2017galaxy, edwards2018pyultralight,
li2019numerical, may2021structure, musoke2024ultradarkjl},
and smoothed particle hydrodynamics (SPH) methods
\citep[e.g.,][]{irsic2017first, nori2018axgadget}.
For a comprehensive review of numerical methods for wave-CDM simulations,
see \cite{schive2026fuzzydark};
for a summary table of existing wave-CDM simulation codes,
see \cite{meng2026wavedmjladaptablea}.

%% July 11, xbdong revised:
In this work, we carry out the simulations using \texttt{WaveDM.jl}\,%
\footnote{\texttt{WaveDM.jl}: \url{https://github.com/JuliaAstroSim/WaveDM.jl}\,.},
a numerical framework for evolving wave dark matter together with ordinary matter in galaxies.
It implements the pseudo-spectral split-step Fourier solver and 
provides tools specially designed for galaxy-scale simulations.
The numerical algorithms, code validation, and quasi-stationarity diagnostics are described in detail 
in the \texttt{WaveDM.jl} code paper \citep{meng2026wavedmjladaptablea}.  
In this section, we summarize the setup specific to the present study: 
the numerical domain and resolution criteria, the treatment of the baryonic component, 
and the construction of the initial wave-CDM halos.
%% xbdong, checklist: only the simulation setup specific to the present study, 
%% : the computational domain, resolution criteria, integration time, 
%% radial range used for the RAR analysis, baryonic model, 
%% and initial conditions for the wave-CDM halos.

\subsection{Simulation setup}
\label{sec:simulations}

The time-dependent SPE is evolved with a second-order split-step Fourier method.
We use computational domains with $256^3$ to $768^3$ grid points, 
depending on the resolution required for each galaxy and boson mass.  
The grid spacing is chosen to satisfy the Nyquist condition $\Delta x \le \lambda_{\mathrm{dB}}/2$.
Resolving the de~Broglie scale is essential for numerical accuracy,
because insufficient resolution at this scale can lead to incorrect dynamics
on larger spatial scales \citep[e.g.,][]{li2019numerical}.

The box size is chosen as $L_x \gtrsim \max(10\lambda_{\mathrm{dB}}, 8r_0)$ 
in order to reduce boundary effects, 
where $r_0$ is the NFW scale radius of the corresponding halo (see below).
We adopt $m_{22}=0.1$ as the baseline boson mass in this work,  %%July14
where $m_{22}\equiv m_a/(10^{-22}\,\mathrm{eV})$.  
For each galaxy, we perform simulations over the range $0.03\le m_{22}\le 0.2$, 
spanning nearly an order of magnitude in boson mass.
For NGC~720, we extend the simulations up to $m_{22}=0.4$.

We choose the timestep according to the accuracy criterion used in previous wave-CDM simulations 
\citep{mocz2017galaxy,may2021structure,may2023halo}, as follows,
\begin{equation}
    \Delta t \le \Delta t_{\max} = 0.5 \min\biggl(\frac{4}{3\pi}\frac{m}{\hbar} \Delta x^{2},\; 2\pi\frac{\hbar}{m}\frac{1}{\max{|\Phi|}}\biggr) \,.
\end{equation}
% For the timestep constraint, the potential term is initially evaluated using the maximum absolute value of $\Phi$ over the computational domain.
We integrate for $3.0$ Gyr,
exceeding the empirically measured virialization time of $\sim 1.5$ Gyr in our simulations.

For consistency with \cite{lelli2017one}, 
we restrict the quantitative analysis to radii satisfying
$\max(2\Delta x, \, 0.1 R_{\mathrm{eff}}) \le r \le 4 R_{\mathrm{eff}}$\,,
where $\Delta x$ is the grid spacing and $R_{\mathrm{eff}}$ is the effective radius.
The $r > 0.1 R_{\mathrm{eff}}$ cut follows \cite{lelli2017one},
and is intended to exclude the innermost regions where possible feedback from active galactic nuclei,
gravitational effects of the central supermassive black hole,
and finite observational resolution may introduce systematic errors.
The $2\Delta x$ cut further ensures adequate numerical resolution
for finite differencing the gravitational potential to obtain accelerations.
The upper bound of $4 R_{\mathrm{eff}}$ is expected to reduce contamination from group-scale potentials
for elliptical galaxies residing in galaxy groups \citep{lelli2017one}.
% \citep[see discussion in][]{lelli2017one}.

\subsection{Baryonic component}\label{sec:baryon_model}

For each galaxy, we adopt the observational baryonic mass distribution as used by \cite{lelli2017one}, 
with the parameters summarized in Sect.~\ref{sec:app_sample}.  
The baryonic component is kept fixed throughout the simulation 
and enters the SPE (Eqs.~\ref{eqn:GPE} and \ref{eqn:Poisson}) only through its static gravitational potential.

In the numerical implementation, this fixed baryonic distribution is represented by an $N$-body particle realization 
rather than by a grid-based density field. 
This has two practical advantages.  
First, the baryonic acceleration $g_\mathrm{b}(r)$ entering the RAR 
can be evaluated directly from the particles, 
avoiding the loss of accuracy associated with finite-difference derivatives of a grid-based potential.  
Second, the vacuum boundary condition for the baryonic potential 
is naturally satisfied by the $N$-body force solver implemented in \texttt{WaveDM.jl} \citep{meng2026wavedmjladaptablea}.  
We apply a Plummer-equivalent force softening of \mbox{$\sim$\,1 kpc}.

For the X-ray ellipticals considered here, 
the hot-gas contribution is subdominant over the radial range used in the RAR analysis 
\citep{milgrom2012testing,lelli2017one}.  
We therefore neglect the hot gas in the baryonic component.

\subsection{Initial conditions for wave-CDM halos}
\label{sec:IC}

We construct the initial conditions (ICs) for the wave-CDM halos as follows.
Because the subsequent evolution is allowed to relax toward a NEQSS, 
the initial density distribution is chosen as a simple top-hat profile 
rather than as a tuned equilibrium configuration:
\begin{equation}
\rho_\mathrm{top-hat}(r) =
\begin{cases}
\rho_\mathrm{ic}, & r \le r_\mathrm{tr}\,, \\
0, & r > r_\mathrm{tr}\,,
\end{cases}
\end{equation}
where $r_\mathrm{tr}$ is a somewhat arbitrary transition radius. 
For each galaxy, we choose $\rho_\mathrm{ic}$ such that 
the initial total DM mass enclosed within the simulation box is comparable to 
that estimated from the corresponding NFW fit to 
the halo profile used by \cite{lelli2017one}
(see Appendix~\ref{sec:app_nfw} for details).
By default, we set $r_\mathrm{tr}=r_0$, where $r_0$ is the scale radius of the NFW halo.
We find that the final results,
such as the best-fit values of $a_0$ and the density profiles of the virialized halos,
are not sensitive to the specific choice of the truncation radius $r_\mathrm{tr}$.
The amplitude of the wave function is then assigned according to
$|\psi|=\sqrt{\rho/m}$.

%% now the IC for the phase of psi:
The initial phase of $\psi$ is constructed by integrating the velocity field,
using the relation $\mathbf{v}=\nabla S$ 
(here in dimensionless form for simplicity, see Eq.~\ref{eqn:fluid_velocity_field}).  
We set the characteristic velocity scale to the local circular velocity,
$v_c(r)=\sqrt{r\,\mathrm{d} \Phi_\mathrm{total}/\mathrm{d} r}$, 
and assign the velocity field as
\begin{equation} \label{eq:vel_rot}
\mathbf{v} = \zeta_{\mathrm{vel}} \left[ \zeta_{\mathrm{rot}} \, v_{c} \, \hat{\mathbf{e}}_\phi
+ (1-\zeta_{\mathrm{rot}}) \, v_{c} \, \hat{\mathbf{e}}_{\mathrm{random}} \right],
\end{equation}
where $\zeta_\mathrm{vel}$ controls the overall velocity amplitude,
$\zeta_\mathrm{rot} \in [0,1]$ specifies the fraction assigned to the coherent azimuthal component, 
$\hat{\mathbf{e}}_\phi$ is the azimuthal unit vector,
and $\hat{\mathbf{e}}_\mathrm{random}$ is a block-wise random unit vector.
Unless otherwise stated, we apply the ``pooling'' strategy, 
i.e., velocity-field coarse graining, introduced in the \texttt{WaveDM.jl} code paper
\citep{meng2026wavedmjladaptablea}.

%% July10, xbdong:
Finally, we perturb the initial $\psi(\mathbf{r})$ 
by adding random noise to the IC data of the three-dimensional density and velocity fields,
in order to accelerate halo relaxation.  
During the subsequent evolution toward a NEQSS (see Section~\ref{sec:virialization}), 
the halo is expected to lose memory of the detailed top-hat initialization.  
As a result, the relaxed density profile will differ substantially from the initial one, 
and the final virialized halo mass will differ from the initial mass enclosed within the simulation box, 
by a somewhat unpredictable amount that depends on the relaxation process.

% \newcolumn

\section{Results}\label{sec:results}

In this section, we present the simulation and data-analysis results 
for the eight X-ray elliptical galaxies.  
% 2026.8.13, after the referee's report (Aug3); adding halo fluctuations, changing the order of paragraphs: 
%%
%% 20260713 R. Y. Meng
%For each boson mass,
%we carry out about 20 simulation runs with various combinations of initial total halo mass and velocity field setup,
%and select the final snapshot that best reproduces the RAR.
For each galaxy and each adopted boson mass, we perform about twenty simulation runs with different initial-condition realizations.
These runs explore a range of initial halo masses, velocity-field setups, etc., as described in Sect.~\ref{sec:numerical}.
From this set of trial realizations,
we select the simulation whose final snapshot provides the closest match to the empirical RAR
over the radial range adopted for the analysis (in the sense of least-squares fitting).
The results presented below therefore correspond, for each (galaxy, boson mass) pair,
to the numerical wave-CDM realization that best reproduces the observed acceleration relation.

%% Aug13, xbdong, a new paragraph:
All data-analysis results, unless otherwise stated, are measured from the final snapshots at $t=3.0\,\mathrm{Gyr}$, 
by which time the simulated halos have evolved into virialized NEQSSs.
	From the perspective of actual observations,
	what astronomers measure for a galaxy is essentially single-epoch,
	which corresponds to the output snapshot of our numerical simulation.
Note that a wave-CDM halo in NEQSS exhibits persistent density fluctuations on a characteristic timescale, 
namely the de~Broglie timescale (see \cite{hui2021wavedark}). 
The central soliton likewise oscillates on a timescale of the same order of magnitude 
\citep[see, e.g.,][]{schive2020soliton,li2021oscillations}. 
This de~Broglie timescale is estimated as
\begin{equation}
    t_\mathrm{dB} \equiv \frac{2\pi \hbar}{m_a v^{2}}\, .
\end{equation}
For the simulated halos, the typical de~Broglie timescales are broadly on the order of $10^7\,\mathrm{yr}$,
which is much shorter than the total simulated time and the NEQSS duration. 
%%%
We illustrate the temporal variation of the best-fit value of $a_0$
for the NGC~720 simulation with $m_{22}=0.1$ in Fig.~\ref{fig:NGC720_a0_time} 
(see Sect.~\ref{sec:app_temporal-variation}).
	Over the NEQSS interval $2 \lesssim t/\mathrm{Gyr} \lesssim 3$,
the temporal mean and standard deviation of the sampled values are
$\left\langle a_0 \right\rangle_t=1.1963\times10^{-10}\,\mathrm{m\,s^{-2}}$
and $\sigma_t(a_0) = 0.0706\times10^{-10}\,\mathrm{m\,s^{-2}}$, respectively.
This shows that the time variation of $a_0$ during the NEQSS remains small ($\sim 5.9\%$ relative dispersion),
so that the value of $a_0$ at the final snapshot ($t=3\,\mathrm{Gyr}$) is representative of the NEQSS. 
%% Sep10 adds:
Such stability of the best-fit $a_0$ values is mainly due to the global nature of the RAR fitting, 
which uses the $g_\mathrm{total}$ and $g_\mathrm{b}$ data at all radii without distinguishing their spatial locations.

% new para, xbdong, July13night:
Here, we must reiterate the scope of the present analysis. 
The primary purpose of this work is to test whether virialized wave-CDM halos \emph{can} reproduce 
the observed RAR for the adopted baryonic mass distributions.
It is not intended as a systematic survey of
the full distribution of possible relations between $g_\mathrm{total}$ and $g_\mathrm{b}$ 
among all viable wave-CDM halo realizations.
Accordingly, our results should be interpreted as 
\emph{demonstrating the existence of wave-CDM realizations compatible with the RAR}, 
rather than as establishing that such agreement is obtained for every choice of initial conditions, 
or that the selected best-matching numerical realizations are typical of the full wave-CDM halo population.

\subsection{Reproducing the RAR with wave-CDM halos} \label{sec:results_RAR}

\begin{figure*}[!ht] %% Figure 1
    \centering
    \includegraphics[width=0.85\textwidth]{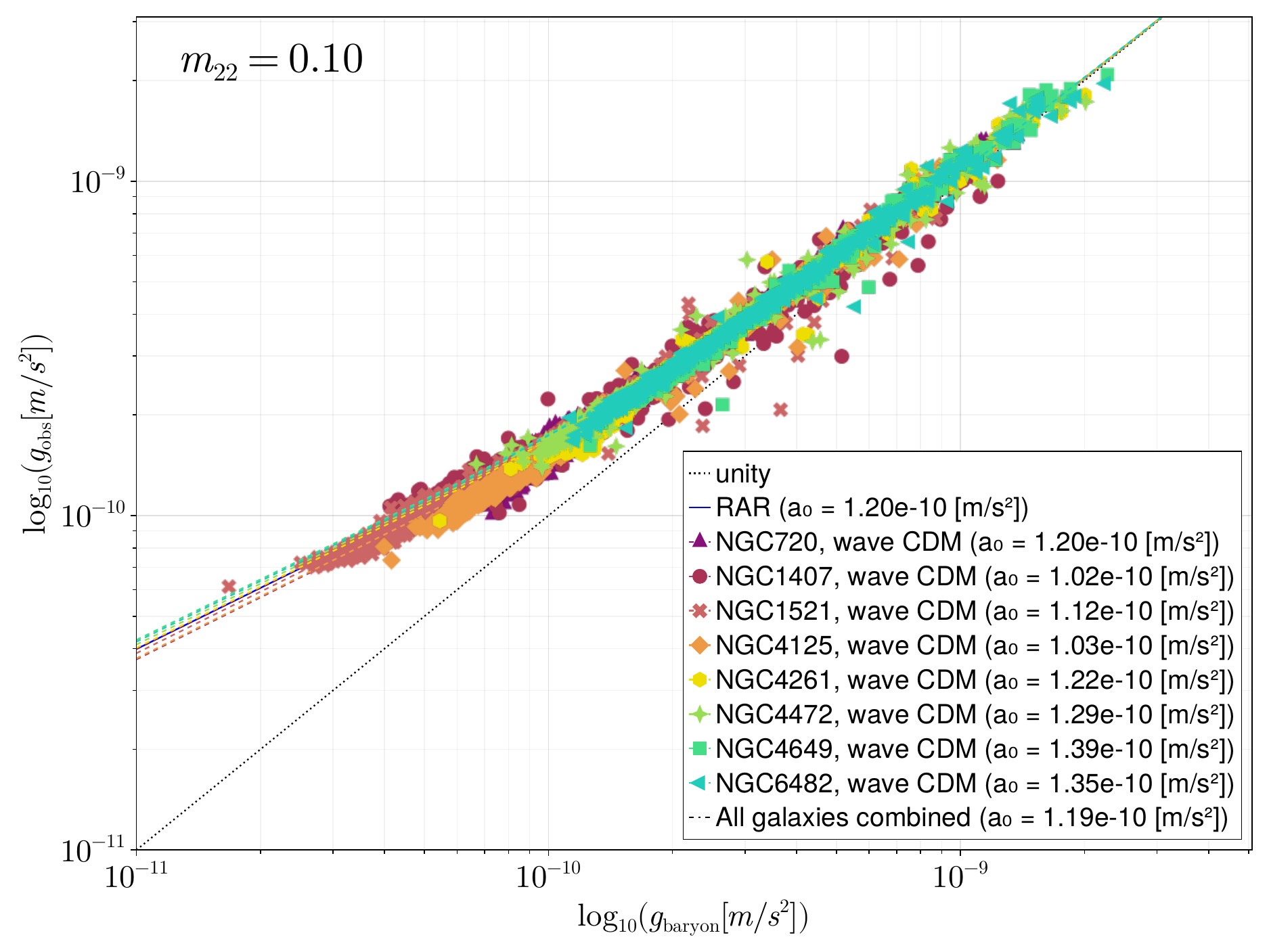}
    \caption{
        Radial acceleration relation for all eight X-ray elliptical galaxies at the final snapshot ($t=3$ Gyr).
        X-axis: baryonic acceleration $g_\mathrm{b}$.
        Y-axis: wave CDM simulated total acceleration, denoted $g_{\mathrm{obs}}$ here for consistency with the RAR literature.
        Solid blue curve: empirical RAR with canonical $a_0=1.20\times10^{-10}\,\mathrm{m\,s^{-2}}$.
        Colored points from the eight galaxies lie close to the empirical RAR.
        Dashed lines show galaxy-specific least-squares fits of Eq.~\ref{eq:RAR};
        their fitted $a_0$ values differ from the canonical value by at most $16\%$.
        The boson mass is $m_{22} = 0.1$ (where $m_{22} \equiv m_a/10^{-22}\,\mathrm{eV}$).
    }
    \label{fig:RAR_all_galaxies}
\end{figure*}

Fig.~\ref{fig:RAR_all_galaxies} presents the simulation results in the RAR plane.  
Each data point shows the total gravitational acceleration measured in the simulated halo
(denoted as $g_\mathrm{obs}$ in the figure)
as a function of the baryonic acceleration $g_\mathrm{b}$ at a given radius.
The best-matching wave-CDM realizations lie close to the empirical RAR
(solid blue curve, with the canonical $a_0 = 1.20 \times 10^{-10} \, \mathrm{m\,s^{-2}}$),
showing that virialized wave-CDM halos \emph{can reproduce} 
the relation between baryonic and total acceleration observed in disk galaxies.

For each realization, and for the combined sample, we determine
$a_0$ by a least-squares fit of the RAR formula of Eq.~\ref{eq:RAR} 
over the radial range $\max(2\Delta x, \, 0.1 R_{\mathrm{eff}}) \le r \le 4 R_{\mathrm{eff}}$.
%% July14: 
The best-fit values of $a_0$ are listed in Table~\ref{table:RAR_fit_statistics}.  
They span $(1.02$--$1.39)\times10^{-10}\,\mathrm{m\,s^{-2}}$, 
and remain within $\lesssim 16\%$ of the canonical RAR value. 
%% stated already: ($a_0 = 1.20 \times 10^{-10} \, \mathrm{m\,s^{-2}}$).  
The quoted uncertainties in $a_0$ are 
simply the statistical measurement errors directly output by the least-squares fitting code
(based on the final snapshot data).  
They do not include the temporal variation of $a_0$ between snapshots,
the scatter among different initial-condition realizations,
or the systematic uncertainty ($\sim 20\,\%$) associated with the stellar mass-to-light ratio $\Upsilon$
\citep{lelli2017one}.
Fig.~\ref{fig:RAR_all_galaxies} also demonstrates the fitting results to individual galaxies (the dashed lines),
which are evidently close to the empirical RAR.

We have also calculated and inspected $g_\mathrm{total}$ and $g_\mathrm{b}$
for the selected best-matching realizations of all eight galaxies at every adopted boson mass.  
The resulting acceleration relations are all similar to that shown in Fig.~\ref{fig:RAR_all_galaxies}.

\begin{table}[H] % Table 1
    \centering
    \footnotesize
    \caption{%\normalsize
        Best-fit $a_0$ for individual galaxies.
        The boson mass is $m_{22} = 0.1$.
        Columns: (1) Galaxy name;
        (2) Best-fit characteristic acceleration scale;
        (3) Number of radial data points.
        Throughout this paper, the quoted uncertainty in $a_0$
        represents only the formal statistical errors from the least-squares fit to the final snapshot data.
        It does not include the temporal variation of $a_0$ between snapshots,
        the scatter among different initial-condition realizations,
        or systematic uncertainties from the stellar mass-to-light ratio
        $\Upsilon$ ($\sim 20\,\%$; \cite{lelli2017one}).
        Individual galaxy $a_0$ values are therefore physically meaningful
        only at the $\sim 20\,\%$ level.
    }
\smallskip
    \begin{tabular}{lcc}
        \hline
        Galaxy & $a_0$ ($10^{-10}\,\mathrm{m\,s^{-2}}$) & $N_{\mathrm{points}}$ \\
        \hline
        NGC 720  & $1.2034 \pm 0.0138$ & 900 \\
        NGC 1407 & $1.0167 \pm 0.0181$ & 1080 \\
        NGC 1521 & $1.1238 \pm 0.0108$ & 1384 \\
        NGC 4125 & $1.0329 \pm 0.0132$ & 1032 \\
        NGC 4261 & $1.2248 \pm 0.0223$ & 676 \\
        NGC 4472 & $1.2898 \pm 0.0265$ & 696 \\
        NGC 4649 & $1.3879 \pm 0.0317$ & 428 \\
        NGC 6482 & $1.3496 \pm 0.0285$ & 528 \\
        \hline
        All ($m_{22}=0.1$) & $1.1903 \pm 0.0070$ & 6724 \\
        \hline
    \end{tabular}
    \label{table:RAR_fit_statistics}
\end{table}

% \newcolumn
% \newpage

\subsection{Insensitivity of $a_0$ to boson mass} \label{sec:results_mass}

\begin{figure}[H]
    \centering
    \includegraphics[width=1.0\columnwidth]{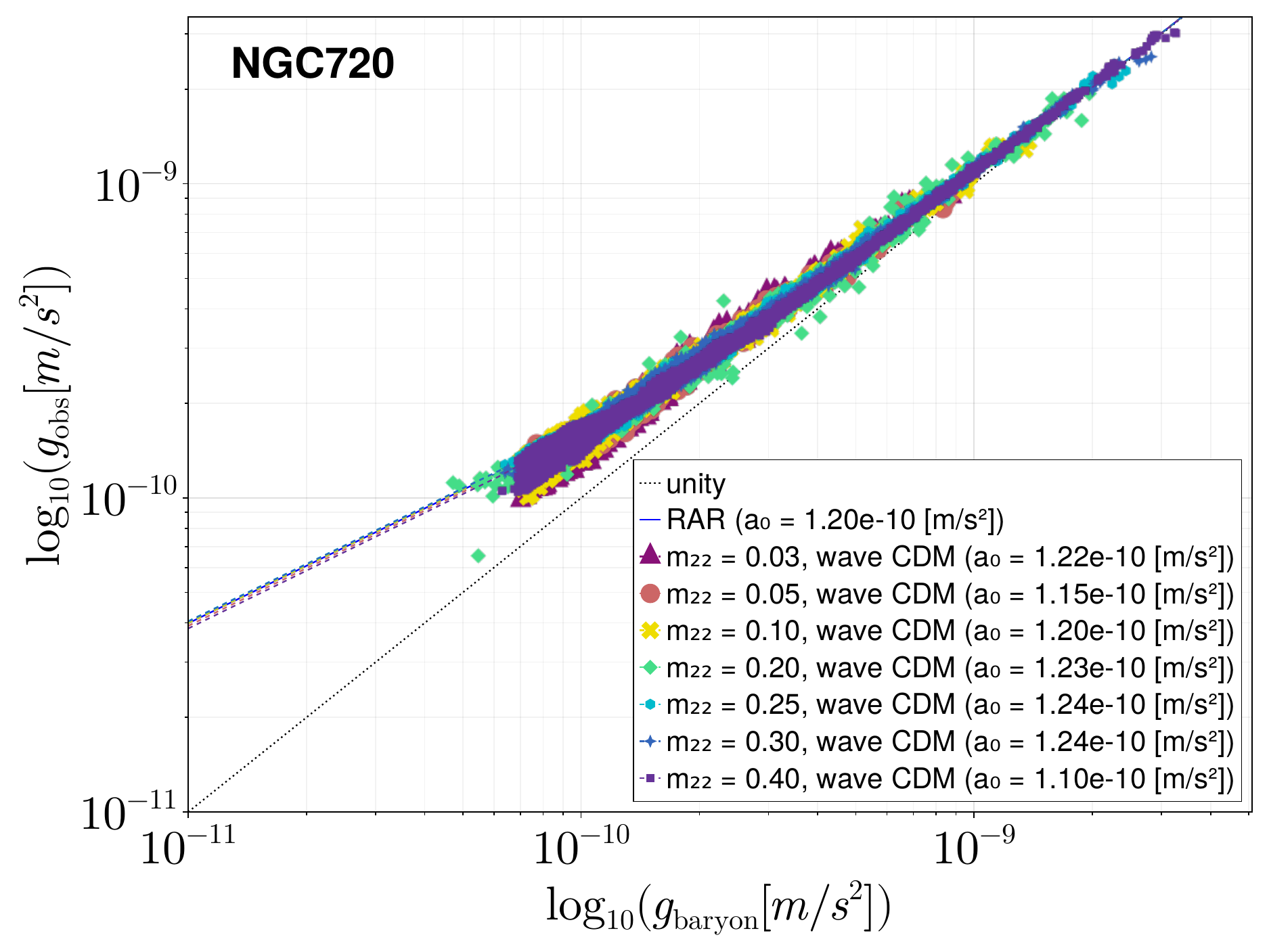}
    \caption{
        Boson-mass insensitivity of the RAR for NGC~720 at the final snapshot ($t=3$ Gyr) for $m_{22}=0.03$--$0.4$.
        X-axis: baryonic acceleration $g_\mathrm{b}$.
        Y-axis: wave CDM simulated total acceleration, denoted $g_{\mathrm{obs}}$ here for consistency with the RAR literature.
        Solid blue curve: empirical RAR with $a_0=1.20\times10^{-10}\,\mathrm{m\,s^{-2}}$.
        Colored points show wave CDM simulations at different boson masses ($m_{22}=0.03$--$0.4$).
        Dashed lines show per-mass least-squares fits of Eq.~\ref{eq:RAR}.
    }
    \label{fig:RAR_NGC720_main}
\end{figure}

Fig.~\ref{fig:RAR_NGC720_main} shows the acceleration relation for the selected realizations of NGC~720 
over the range $0.03\le m_{22}\le 0.4$.  
The acceleration data points, sampled at different radii, 
maintain a tight correlation between $g_\mathrm{b}$ and $g_\mathrm{total}$ for all tested values of $m_{22}$.  
Thus, wave-CDM halos with suitable ICs can reproduce the observed RAR throughout this boson-mass range.  
%%%
The mass-specific fits (dashed curves) are nearly indistinguishable from the canonical RAR, 
and their best-fit values of $a_0$ remain close to the canonical value (Table~\ref{table:mass_dependence}).

Importantly, Table~\ref{table:mass_dependence} further shows that 
the variation in the best-fit $a_0$ exhibits no systematic trend with $m_{22}$.

Likewise, we have performed fits to the $g_\mathrm{total}$--$g_\mathrm{b}$ relation 
for the other seven galaxies at each adopted boson mass, 
as well as for the combined $m_{22}$-specific samples of the best-matching simulations 
for all eight galaxies at different boson masses.
In all cases, the best-fit values of $a_0$ remain close to the canonical value 
and show no systematic dependence on $m_{22}$.  
The fitting results for the four combined samples at different boson masses 
are listed in the Appendix (see Table~\ref{table:combined_RAR_all}).

\begin{figure}[H]
    \centering
    \includegraphics[width=1.0\columnwidth]{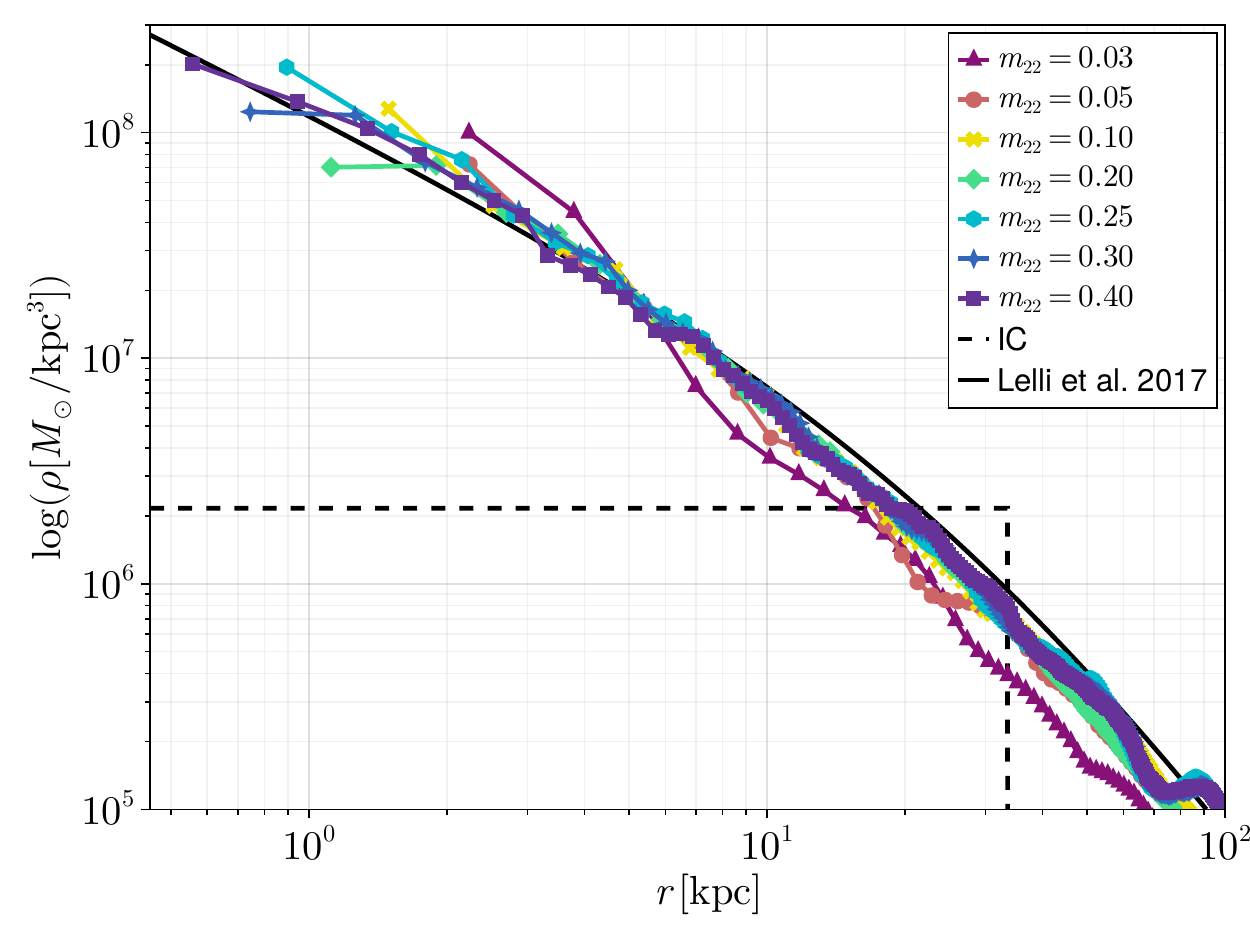}
    \caption{
	    Density profiles of the simulated wave-CDM halos of NGC~720 for different boson masses.
	    The black dashed curve shows the initial top-hat profile, 
	    while the colored dashed curves show the virialized wave-CDM profiles at $t=3.0\,\mathrm{Gyr}$.
	    For comparison, the black solid curve shows the NFW halo profile 
	    inferred from the Lelli \etal\ data \citep{lelli2017one} (see Appendix~\ref{sec:app_nfw}).
    }
    \label{fig:NGC720_res_profile}
\end{figure}

It is also instructive to examine the radial density profiles of the selected best-matching NGC~720 realizations 
at different boson masses, as shown in Fig.~\ref{fig:NGC720_res_profile}.  
As $m_{22}$ decreases from $0.4$ to $0.03$, 
the mass-specific profiles differ to a certain degree on kpc and sub-kpc scales, 
plausibly associated with the corresponding de~Broglie wavelengths, 
but show no clear systematic trend overall.  
At $r\gtrsim10\,\mathrm{kpc}$, their slopes are broadly similar and approximately parallel to that of the NFW profile.
A detailed investigation of the density profiles is beyond the scope of this work.

\begin{table}[H] % Table 2
% \begin{table}[!tb] % Table 2
    \centering
    % \footnotesize
    \caption{%\normalsize
        Best-fit $a_0$ at different boson masses for NGC~720.
        Columns: (1) Boson mass in units of $10^{-22}$ eV;
        (2) Best-fit characteristic acceleration $a_0$ and the $\pm1\sigma$ formal statistical uncertainty 
        returned by the least-squares fitting code;
        (3) Number of radial data points used in the fit.
    }
\smallskip
    \renewcommand{\arraystretch}{1.0}
    \begin{tabular}{ccc}
        \hline
        $m_{22}$ & $a_0$ ($10^{-10}\,\mathrm{m\,s^{-2}}$) & $N_{\mathrm{points}}$ \\
        \hline
        0.03 & $1.2152 \pm 0.0305$ & 388 \\
        0.05 & $1.1475 \pm 0.0220$ & 388 \\
        0.10 & $1.2034 \pm 0.0138$ & 900 \\
        0.20 & $1.2301 \pm 0.0128$ & 1636 \\
        0.25 & $1.2443 \pm 0.0077$ & 2556 \\
        0.30 & $1.2371 \pm 0.0060$ & 3668 \\
        0.40 & $1.0972 \pm 0.0038$ & 6548 \\
        \hline
    \end{tabular}
    \label{table:mass_dependence}
\end{table}

% \begin{figure*}[!htbp]

% \begin{figure*}[!htbp]

\newcolumn

\subsection{Physical origin of RAR's insensitivity to boson mass}
\label{sec:results_reason}

The insensitivity of the RAR to boson mass can be traced to a simple but important property of the quantum-pressure term 
in the quantum Euler equation (Eq.~\ref{eqn:Mad_Euler}). 
For a power-law DM density profile, $\rho(r) \propto r^{-n}$, the radial derivative of the quantum potential scales as
\begin{equation}
\label{eqn:quantum-pressure-for-NFW}
\grad Q
= \frac{\partial Q}{\partial r}
= \frac{\hbar^2}{4 m_a}\, n\, (n-2)\, r^{-3} \, .
\end{equation}
This expression highlights two relevant facts. 
First, the quantum-pressure force decreases rapidly with radius, $\propto r^{-3}$. 
Second, for all realistic dark-matter halos, whose density profiles are approximately NFW-like over a broad radial range, 
the local logarithmic slope is close to $n \simeq 2$. 
In this regime, the prefactor $n(n-2)$ strongly suppresses the quantum-pressure contribution.

%% new para:
The boson mass enters the quantum Euler equation only through the quantum-pressure term. 
Therefore, over the radial range in which the RAR is probed, 
the influence of boson mass is strongly suppressed due to the above two simple facts.

%%new para: check with the simulation data:
As described in Sect.~\ref{sec:virialization}, the relaxed halo settles into a NEQSS
in global virialization (see Fig.~\ref{fig:app_energy})
in which the angle-integrated density profile $\rho(r)$ is stationary (see Figs.~\ref{fig:app_mass_profile} and \ref{fig:app_radii}) 
and the angle-averaged velocity field satisfies 
a Jeans-like equation in the radial direction (see Eq.~\ref{eq:Jeans} 
and the bottom panel of Fig.~\ref{fig:NGC720_vel_field}).
The velocity moments and gravitational potential are measured from the simulation data 
(see \cite{meng2026wavedmjladaptablea} for details). %% July11.
Visualizations of the velocity field 
for the NGC~720 halo are provided in the Appendix (Fig.~\ref{fig:NGC720_vel_field}, top panel) 
for the interested reader.

% new para: July11, xbdong:
We assess the importance of the quantum-pressure term in Eq.~\ref{eq:Jeans} with the simulation data in the following way. 
Because evaluating \(\mathrm{d}Q/\mathrm{d}r\) directly from noisy simulation profiles 
requires numerical differentiation and is therefore not robust, 
we do not inspect this term directly.
Instead, the bottom panel of Fig.~\ref{fig:NGC720_vel_field} 
compares the velocity dispersion measured from the simulation data 
with $\sigma_\mathrm{Jeans}$, which is obtained by integrating Eq.~\ref{eq:Jeans} 
under the assumptions of isotropic velocity dispersion and negligible quantum pressure.  
%% July 12, xbdong:
The approximate agreement between 
the measured radial, polar, and azimuthal velocity dispersions and the $\sigma_\mathrm{Jeans}$ curve 
over the radial range used in the RAR analysis
suggests that no large residual support from quantum pressure is required there.

%% reviewer #4: cosmological boson mass
	We should note that this study explores only a limited range of boson masses. 
In particular, the values considered around $m_a \sim 10^{-24}$--$10^{-23}\,\mathrm{eV}$ 
lie well below the $\sim 10^{-21}\,\mathrm{eV}$ scale commonly examined in cosmological analyses. 
We focus on this low-mass regime because increasing $m_a$ 
shifts the characteristic de~Broglie length and time scales to smaller values, 
making the resulting wave structures progressively more difficult to resolve in our simulations. 
The halos would approach the behavior expected for particle CDM, 
as the wave effects become less pronounced \emph{at fixed numerical resolution}. 
Extending the analysis to substantially larger boson masses would therefore 
require significantly higher spatial resolution, smaller time steps, and greater computational cost, 
and is deferred to future work.

%% \bigskip
Finally, we emphasize again a limitation of the present work. 
Our simulations demonstrate that wave CDM is compatible with the observed radial acceleration relation in X-ray elliptical galaxies, 
and that the existence of this tight empirical relation does not by itself rule out wave CDM. 
However, the present study does not explain the origin of the tight baryon--dark matter correlation \textit{per se}. 
In particular, it does not identify a fundamental mechanism 
that explains why the spatial distribution and gravitational field of dark matter 
should appear to be so closely linked to those of ordinary matter. 
Clarifying this origin remains an open problem.

\section{Conclusions}\label{sec:conclude}

We have presented three-dimensional Schr\"odinger--Poisson simulations of wave-CDM halos for eight X-ray elliptical galaxies
from the compilation of \cite{lelli2017one}.
We examined whether the virialized wave-CDM halos can reproduce the observed radial acceleration relation (RAR),
and obtained two main findings.
\begin{itemize}
\item  \textbf{Wave CDM can reproduce the RAR in elliptical galaxies.}
For the baseline boson mass \mbox{$m_{22} \equiv m_a / 10^{-22}\mathrm{eV} = 0.1$},
the combined fit to $6724$ radial data points from the selected best-matching numerical realizations of all eight galaxies
yields $a_0 = 1.19 \times 10^{-10}\,\mathrm{m\,s^{-2}}$,
close to the canonical value $1.20 \times 10^{-10}\,\mathrm{m\,s^{-2}}$
reported for late-type disk galaxies \citep{lelli2017one}.
The $a_0$ values from the fitting of individual galaxies 
span $(1.02$--$1.39) \times 10^{-10}\,\mathrm{m\,s^{-2}}$
(Table~\ref{table:RAR_fit_statistics}),
deviating from the canonical value by at most $\sim 16\,\%$ ---
smaller than the $\sim 20\,\%$ systematic uncertainty
from the stellar mass-to-light ratio $\Upsilon$ \citep{lelli2017one}.
Each relaxed system of these realizations
exhibits a tight correlation between $g_\mathrm{b}$ and $g_{\mathrm{total}}$,
with residual scatter similar to that reported for late-type galaxies.
Thus, these virialized halos of collisionless wave CDM 
\emph{can reproduce} the RAR without invoking outflow feedback (from supernovae or active galactic nuclei), finite-temperature effects, 
dark matter self-interactions, or any coupling mechanism between dark matter and baryons.

\item  \textbf{The RAR is insensitive to boson mass $m_a$, and the characteristic acceleration $a_0$ does not correlate with $m_a$.}
Over the range $0.03 \le m_{22} \le 0.2$, for which all eight galaxies were simulated,
we find tight correlations between baryonic and total gravitational accelerations in all cases.
%% $g_\mathrm{b}$ and $g_{\mathrm{obs}}$ in all cases.
The best-fit characteristic acceleration ($a_0$) varies only from $(1.06$--$1.19) \times 10^{-10}\,\mathrm{m\,s^{-2}}$
(Table~\ref{table:combined_RAR_all}),
deviating from the canonical value 
$1.20 \times 10^{-10}\,\mathrm{m\,s^{-2}}$ by $\sim 11\,\%$.
For NGC~720, our higher-resolution simulations extend the explored boson mass to $m_{22}=0.4$,
and the resulting value of $a_0$ remains within the same range.
Furthermore, as evidently seen from Fig.~\ref{fig:RAR_NGC720_main} and Fig.~\ref{fig:NGC720_res_profile}, 
the variation of $a_0$ shows no systematic trend with $m_a$.

\end{itemize}

\medskip

As analyzed in Sect.~\ref{sec:results_reason}, the physical origin of this insensitivity is simple and transparent. 
In the Madelung formulation, the boson mass enters the quantum Euler equation only through the quantum-pressure term. 
However, at the radii relevant for the RAR, the quantum-pressure force is strongly suppressed: 
it decreases rapidly with radius,
and is further reduced when the halo density profile is close to an $r^{-2}$ form. 
Consequently, varying $m_a$ has little effect on the force balance in the radial range probed by the RAR.

Instead of being counter-balanced by the boson-mass-dependent quantum pressure,
in relaxed wave-CDM halos that have settled into a non-equilibrium quasi-stationary state (NEQSS) sustained by weak wave turbulence, 
the gravitational field at RAR-relevant radii is dynamically balanced primarily by the halo velocity field, 
including velocity-dispersion pressure and possible bulk motions such as rotation. 
This point seems relevant to all rotation-curve based analyses of wave CDM: 
unless the observations directly probe the quantum-pressure-dominated region, 
rotation curves alone are not expected to provide a robust constraint on the boson mass. 
We will examine this issue in more detail in forthcoming work.

In fact, this insensitivity to particle mass is the essence of the CDM paradigm generically 
(see \cite{dong2025promise} for more discussions and the implication concerning the recent puzzles from JWST). 
Most observational quantities are measured on macroscopic scales, 
where the system is effectively coarse-grained over the kinetic scales and corresponding nonlinear kinetic processes
(let alone the microscopic scale and dynamical processes).
Thus the detailed microphysical identity of the dark matter particle is largely hidden. 
Whether the dark sector consists of particles or ultralight fields, fermions or bosons, 
its phenomenology can remain effectively the generic trait of collisionless CDM, 
as long as quantum or other microphysical effects are dynamically subdominant. 
This separation between microscopic physics and macroscopic behavior 
is one of the reasons for the long phenomenological success of the CDM paradigm.

%%%%%%%%%%%%%%%%%%%%%%%%%%%%%%%%%%%%%%%%%%%%%%%%%%%%%%%
%%% Acknowledgements
%%%%%%%%%%%%%%%%%%%%%%%%%%%%%%%%%%%%%%%%%%%%%%%%%%%%%%%

\section*{Acknowledgements}
We thank the anonymous referee for the constructive comments, which improved this paper.
%%}
This work is supported by the National Natural Science Foundation of China (grant No.~12373013).
We thank Pengfei~Li for helpful comments,
Zhe~Sun and Tao~Tu for years of discussions on ``NingJuTi'' (Bose condensates) and those ``super-'' phenomena,
Wen-Juan Liu and Qian Long for providing computational resources.
We also acknowledge the inspiration of Haofeng Xu's words,
``Hold fast to your original aspiration'' (\textit{A~Daoist Descends the Mountain}).

\section*{Conflict of interest}
The authors declare that they have no conflict of interest.

%%%%%%%%%%%%%%%%%%%%%%%%%%%%%%%%%%%%%%%%%%%%%%%%%%%%%%%
%%% Supplements. ????????, ????
%%%%%%%%%%%%%%%%%%%%%%%%%%%%%%%%%%%%%%%%%%%%%%%%%%%%%%%
%%%%%%%%%%%%%%%%%%%%%%%%%%%%%%%%%%%%%%%%%%%%%%%%%%%%%%%%
%%% Reference section. ?ο?????
%%% citation in the content using "some words~\cite{1,2}".
%%% ~ is needed to make the reference number is on the same line with the word before it.
%%%%%%%%%%%%%%%%%%%%%%%%%%%%%%%%%%%%%%%%%%%%%%%%%%%%%%%

\bibliographystyle{unsrt}
\bibliography{WaveDM_RAR_ETGs}

% \clearpage
% \newpage

\appendix
\setcounter{section}{0}
\renewcommand{\thesection}{\Alph{section}}

\section{Appendix}

\setcounter{figure}{0}
\setcounter{table}{0}

%% \subsection{Sample selection} %July10
\subsection{The eight simulated galaxies, their baryonic and DM components}

The RAR compilation of \cite{lelli2017one} includes nine X-ray luminous early-type galaxies 
with mass profiles inferred from \textit{Chandra} and \textit{XMM} observations
\citep{humphrey2006chandra, humphrey2008lowmass, humphrey2009hydrostatic, humphrey2011census, humphrey2012elixr}.
In this work we use the eight elliptical galaxies,
namely NGC~720, NGC~1407, NGC~1521, NGC~4125, NGC~4261, NGC~4472, NGC~4649, and NGC~6482.  
The remaining system, NGC~1332, is an S0 galaxy with a prominent disk in addition to its spheroidal bulge, 
and thus does not meet our requirement of being structurally simple; 
we therefore exclude it from the present analysis.
Two of these galaxies, NGC~720 and NGC~1521, were previously studied by \cite{milgrom2012testing}
in the context of MOND (modified Newtonian dynamics), 
also focusing on the connection between the observed gravitational field and that expected from the baryonic mass distribution.

% new para, July10:
The eight galaxies are morphologically simple X-ray ellipticals,
for which a spherical approximation is adequate for the purposes of the present simulations.  
In the following two subsections we describe the two ingredients used to construct the galaxy--halo models.  
%% First, we infer a DM profile for each galaxy by subtracting the baryonic contribution
%%  from the total gravitating mass profile compiled by Lelli \etal~\citep{lelli2017one}, 
%%  and fit the result with an NFW profile.
%% Second, we specify the baryonic component using the stellar parameters adopted by Lelli \etal~\citep{lelli2017one}, 
%%  approximating the stellar distribution by a spherical S\'ersic model.

\subsubsection{NFW fits to Lelli \etal's DM halos of the eight galaxies} \label{sec:app_nfw}

The total gravitating mass profiles of the eight galaxies, corresponding to
the $g_\mathrm{obs}(r)$ profiles, 
are taken from \cite{lelli2017one}.  
For each galaxy, we subtract the baryonic contribution from the total-mass profile 
and fit the resulting DM profile with the standard NFW form
\begin{equation}
    \rho_{\mathrm{NFW}}(r) = \frac{\rho_0}{(r/r_0)(1 + r/r_0)^2} \,,
\end{equation}
where $\rho_0$ is the characteristic density and $r_0$ is the scale radius.
The best-fit values of $r_0$ and $\rho_0$ are listed in Table~\ref{table:nfw_parameters}.

%% July 10, xbdong:
These NFW profiles are used in the main text as reference profiles 
for comparison with the simulated wave-CDM density profiles (see, e.g., Figure~\ref{fig:NGC720_res_profile}).  
They are also used to estimate the initial DM mass enclosed within the simulation box for each galaxy.

% \begin{table}[!tb] %Table A1
\begin{table}[H]
    \centering
    % \footnotesize
    \caption{%\normalsize
       NFW halo parameters for the eight elliptical galaxies.
       The scale radius $r_0$ and characteristic density $\rho_0$ are 
	    obtained by fitting the dark-matter mass profiles inferred from \cite{lelli2017one}, 
	    after subtracting the baryonic contribution from the observational total-mass profiles.
    }
\smallskip
    \begin{tabular}{lcc}
        \hline
        Galaxy & $r_0$ & $\rho_0$ \\
               & (kpc) & ($10^6 M_\odot$/kpc$^3$) \\
        \hline
        NGC 720  & 33.6 & 3.73 \\
        NGC 1407 & 53.6 & 2.24 \\
        NGC 1521 & 60.9 & 1.75 \\
        NGC 4125 & 39.7 & 3.05 \\
        NGC 4261 & 38.9 & 3.36 \\
        NGC 4472 & 44.4 & 2.94 \\
        NGC 4649 & 38.8 & 3.41 \\
        NGC 6482 & 42.9 & 3.09 \\
        \hline
    \end{tabular}
    \label{table:nfw_parameters}
\end{table}

\subsubsection{Baryonic model}
\label{sec:app_sample}

The baryonic component in our simulations is specified using the stellar parameters compiled by \cite{lelli2017one}, 
as summarized in Table~\ref{table:baryonic_properties}.  
Following their treatment of the X-ray ellipticals, we model the stellar distribution as spherical.  
For the numerical implementation, the stellar density profile is represented by a S\'ersic model with fixed index $n=4$, 
using the effective radius $R_\mathrm{eff}$ and the $3.6\,\mu\mathrm{m}$ stellar mass-to-light ratio
$\Upsilon_{[3.6]}=0.9\,M_\odot/L_\odot$ adopted by \cite{lelli2017one}.  
The systematic uncertainty in $\Upsilon_{[3.6]}$ induces an intrinsic scatter of 
approximately $20\%$ in the fitted value of $a_0$ \citep{lelli2017one}; 
accordingly, the best-fit $a_0$ values for individual galaxies should be interpreted with this level of systematic uncertainty.

% \begin{table}[!tb] %Table A2
\begin{table}[H]
	\centering
	\caption{%\normalsize
		Parameters of the stellar component adopted for the eight elliptical galaxies.
		The effective radius $R_\mathrm{eff}$, total $3.6\,\mu\mathrm{m}$ luminosity $L_{[3.6]}$, 
		and stellar mass-to-light ratio $\Upsilon_{[3.6]}$ are
		taken from \cite{lelli2017one}.
	}
\smallskip
    \begin{tabular}{lccc}
        \hline
        Galaxy & $R_{\mathrm{eff}}$ & $L_{[3.6]}$ & $\Upsilon_{[3.6]}$ \\
               & (kpc) & ($10^9 L_\odot$) & ($M_\odot/L_\odot$) \\
        \hline
        NGC 720  & 4.5 & 234.5 & 0.9 \\
        NGC 1407 & 7.8 & 540.1 & 0.9 \\
        NGC 1521 & 10.1 & 532.6 & 0.9 \\
        NGC 4125 & 5.7 & 303.1 & 0.9 \\
        NGC 4261 & 4.5 & 345.4 & 0.9 \\
        NGC 4472 & 5.2 & 447.6 & 0.9 \\
        NGC 4649 & 3.6 & 345.7 & 0.9 \\
        NGC 6482 & 4.4 & 427.9 & 0.9 \\
        \hline
    \end{tabular}
    \label{table:baryonic_properties}
\end{table}

%---------------------------------

\subsection{Simulation parameters for NGC 720} \label{sec:app_720}

The numerical parameters adopted for the NGC~720 simulations at different boson masses 
are summarized in Table~\ref{table:NGC720_sim_params}.  
These simulation runs use the same physical box size and total integration time, 
while the grid resolution and time stepping are adjusted 
to better resolve the shorter de~Broglie wavelengths associated with larger boson masses.

\begin{table*}[!tb] %Table A3
% \begin{table*}[!h]
	\centering
	%\footnotesize
	\caption{%\normalsize
		Numerical parameters for the NGC~720 wave-CDM simulations.
		The columns list  
		(1) the boson mass $m_{22} \equiv m_a/(10^{-22}\,\mathrm{eV})$, 
		(2) the grid size $N_x$,
		(3) the box size $L$, 
		(4) the number of time steps $N_t$, 
		(5) the total integration time $T_{\max}$, 
		(6) the number of baryonic particles $N_p$, 
		(7) the boundary absorption coefficient $\alpha$, 
		(8) the pooling-kernel size  (in grid cells), 
		(9) the initial velocity-amplitude factor $\zeta_\mathrm{vel}$, 
		(10) the initial rotation fraction $\zeta_\mathrm{rot}$,
		(11) the grid spacing $\Delta x$, 
		(12) the box size in units of the top-hat transition radius $r_0$, 
		(13) the box size in units of the de~Broglie wavelength $\lambda_\mathrm{dB}$, 
		(14) the number of grid cells per de~Broglie wavelength, 
		and (15) the total mass (enclosed within five scale radii) of the virialized wave-CDM halo at the final snapshot.
	}
    \smallskip
	\resizebox{0.9\textwidth}{!}{
		\begin{tabular}{|c|c|c|c|c|c|c|c|c|c|c|c|c|c|c|}
			\hline $m_{22}$ & $N_x$ & $L$   & $N_t$ & $T_{\max}$ & $N_p$ & $\alpha$ & pooling & $\zeta_\mathrm{vel}$ & $\zeta_\mathrm{rot}$ & $\Delta x$ & $L/r_0$ & $L/\lambda_\mathrm{dB}$ & $\lambda_\mathrm{dB}/\Delta x$ & $M_{\mathrm{total}}(<5r_0)$ \\
			                &       & (kpc) &       & (Gyr)      &       &          &         &                      &                      & (kpc)      &         &                  &                         & ($10^{12} M_\odot$)  \\
			      (1)       &  (2)  & (3)    &  (4) & (5)        &  (6)  & (7)      &  (8)    &  (9)                 &   (10)               & (11)       &  (12)   &  (13)            & (14)                    & (15)  \\                        
			\hline 0.03     & 256   & 403.0 & 1371  & 3.0        & 5000  & 10       & 8       & 0.7                  & 0.9                  & 1.574      & 12.00   & 15.72            & 16.29                   & 1.11                 \\
			\hline 0.05     & 256   & 403.0 &  823  & 3.0        & 5000  & 10       & 8       & 0.7                  & 0.9                  & 1.574      & 12.00   & 26.19            & 9.774                   & 1.26                  \\
			\hline 0.10     & 384   & 403.0 & 1110  & 3.0        & 5000  & 10       & 16      & 0.7                  & 0.9                  & 1.049      & 12.00   & 52.38            & 7.330                   & 1.31                 \\
			\hline 0.20     & 512   & 403.0 & 2570  & 3.0        & 5000  & 10       & 4       & 0.7                  & 0.9                  & 0.787      & 12.00   & 104.8            & 4.887                   & 1.29                 \\
			\hline 0.25     & 640   & 403.0 & 3037  & 3.0        & 5000  & 10       & 32      & 0.7                  & 0.9                  & 0.630      & 12.00   & 131.0            & 4.887                   & 1.32                 \\
			\hline 0.30     & 768   & 403.0 & 3723  & 3.0        & 5000  & 10       & 32      & 0.7                  & 0.9                  & 0.525      & 12.00   & 157.2            & 4.887                   & 1.31                 \\
			\hline 0.40     & 1024  & 403.0 & 4929  & 3.0        & 5000  & 10       & 32      & 0.7                  & 0.9                  & 0.394      & 12.00   & 209.6            & 4.887                   & 1.31                 \\
			\hline
		\end{tabular}
	} %scriptsize , \small
	\label{table:NGC720_sim_params}
\end{table*}

\subsection{Virialization diagnostic} \label{sec:app_virial}

In this section,
we demonstrate virialization diagnostics using a simulation run of NGC~720 for $m_{22}=0.1$.
A detailed description of these diagnostics is provided in the \texttt{WaveDM.jl} code paper \citep{meng2026wavedmjladaptablea}.

\subsubsection{Evolution of overall energy terms}

\begin{figure*}[!htbp]
    \centering
    \includegraphics[width=1.0\textwidth]{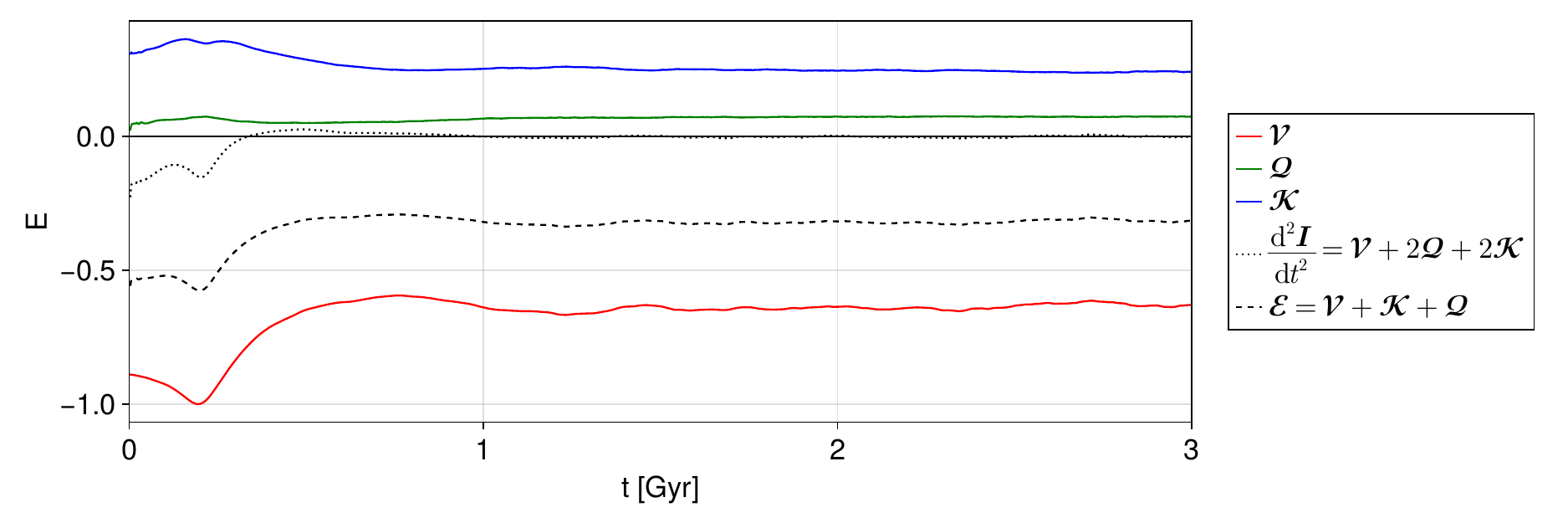}
    \caption{
        Time evolution of overall energy terms of NGC~720.
        Red: gravitational potential energy $\mathcal{V}$;
        green: quantum gradient energy $\mathcal{Q}$;
        blue: kinetic energy $\mathcal{K}$;
        dotted: second time derivative of the moment of inertia
            $\mathrm{d}^2\mathbf{I}/\mathrm{d}t^2 = 2\mathcal{V} + 2\mathcal{Q} + 2\mathcal{K}$
            (the right-hand side of the virial theorem);
        dashed: total energy $\mathcal{E} = \mathcal{V} + \mathcal{K} + \mathcal{Q}$.
        All quantities are normalized to $|\mathcal{V}|$.
        The halo becomes virialized by $\sim 1\,\mathrm{Gyr}$,
        after which $\mathrm{d}^2\mathbf{I}/\mathrm{d}t^2$ oscillates with small amplitude about zero.
        The boson mass is $m_{22}=0.1$.
    }
    \label{fig:app_energy}
\end{figure*}

Fig.~\ref{fig:app_energy} shows the time evolution of the overall energy terms of the NGC~720 halo,
including
the kinetic energy $\mathcal{K}$,
the quantum gradient energy $\mathcal{Q}$,
the gravitational potential energy $\mathcal{V}$,
their sum $\mathcal{E} = \mathcal{V} + \mathcal{K} + \mathcal{Q}$,
and the second time derivative of the moment of inertia $\mathrm{d}^2 \mathbf{I} / \mathrm{d}t^2$.
The halo is virialized once $\mathrm{d}^2 \mathbf{I}/\mathrm{d}t^2$ settles to a small residual oscillation about zero.
Definitions and numerical evaluation of these quantities follow \cite{meng2026wavedmjladaptablea}.
For this run the halo reaches such a virialized state by $\sim 1\,\mathrm{Gyr}$,
and remains in it for the remainder of the $3\,\mathrm{Gyr}$ integration.

\subsubsection{Evolution of mass-fraction radii}

\begin{figure*}[!htbp]
    \centering
    \includegraphics[width=1.0\textwidth]{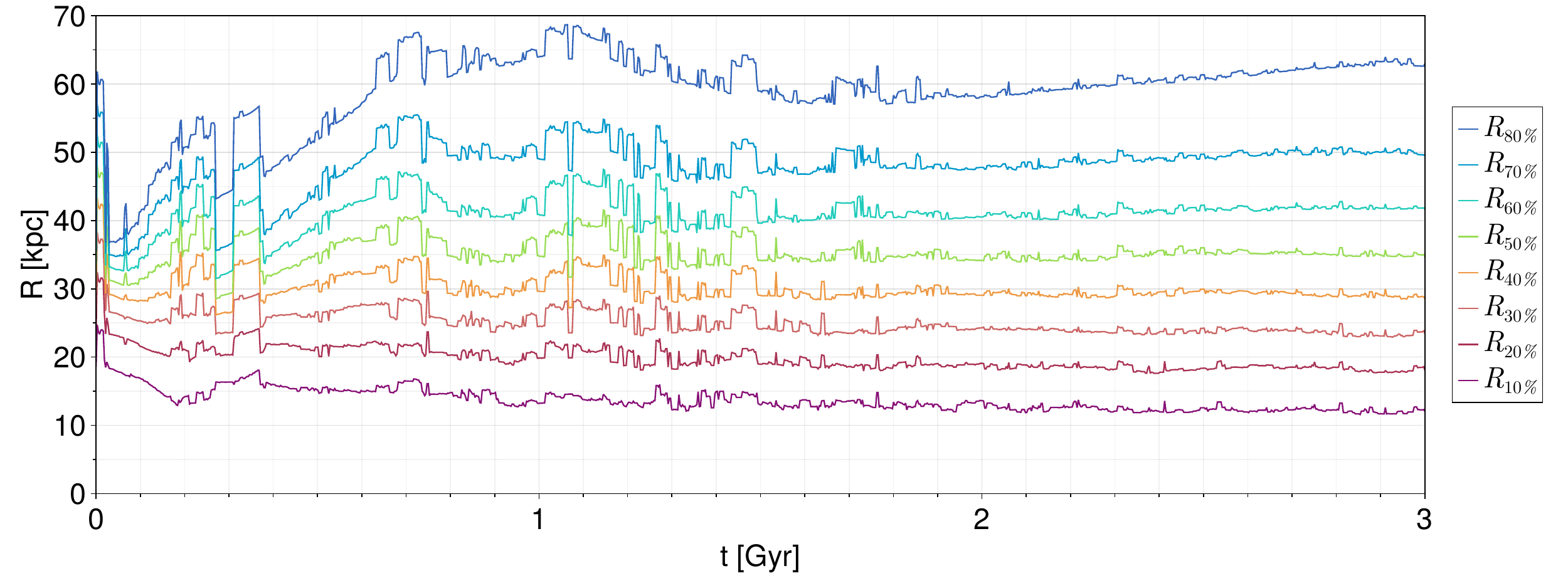}
    \caption{
        Time evolution of mass-fraction radii of NGC~720.
        Each curve tracks the radius enclosing a fixed mass fraction (10 \% to 80 \%, bottom to top).
        The radii reach stationary values by $\sim 2\,\mathrm{Gyr}$,
        indicating that the mass distribution has settled into a quasi-stationary state.
        The boson mass is $m_{22} = 0.1$.
    }
    \label{fig:app_radii}
\end{figure*}

As a complementary, geometry-based diagnostic,
Fig.~\ref{fig:app_radii} tracks the mass-fraction radii of the wave-CDM component (10--80\,\%, bottom to top) over time.
The settling of these radii to nearly constant values by $\sim 2\,\mathrm{Gyr}$ indicates that
the radial mass distribution has reached a stationary state.

\subsubsection{Kinematic diagnostics}

The angle-averaged Jeans equation in the radial direction for virialized wave-CDM halos reads
\begin{equation} \label{eq:Jeans}
    \frac{\mathrm{d}(\rho \sigma_{\mathrm{r}}^{2})}{\mathrm{d} r}+2\frac{\beta(\rho \sigma_{\mathrm{r}}^{2})}{r} - \rho \frac{\langle v_\theta \rangle^2 + \langle v_\phi \rangle^2}{r}=-\rho \frac{\mathrm{d}\Phi_{\mathrm{total}}}{\mathrm{d} r} - \rho \frac{\mathrm{d}Q}{\mathrm{d} r},
\end{equation}
where $\beta=1-(\sigma_{\theta}^{2} + \sigma_{\phi}^{2})/2\sigma_{\mathrm{r}}^{2}$ measures velocity anisotropy,
$Q$ denotes the quantum potential (Eq.~\ref{eqn:Potential_Q}),
and $\sigma_r$\,, $\sigma_\theta$\,, $\sigma_\phi$ are the radial, zenithal,
and azimuthal velocity dispersions.
In deriving this equation, the angle-averaged, mean radial velocity $\langle v_r \rangle$ of the virialized halos
is assumed to be negligible, and thus there is no time dependent term (see \cite{meng2026wavedmjladaptablea}).

\begin{figure*}[!htbp]
    \centering
    \includegraphics[width=0.8\textwidth]{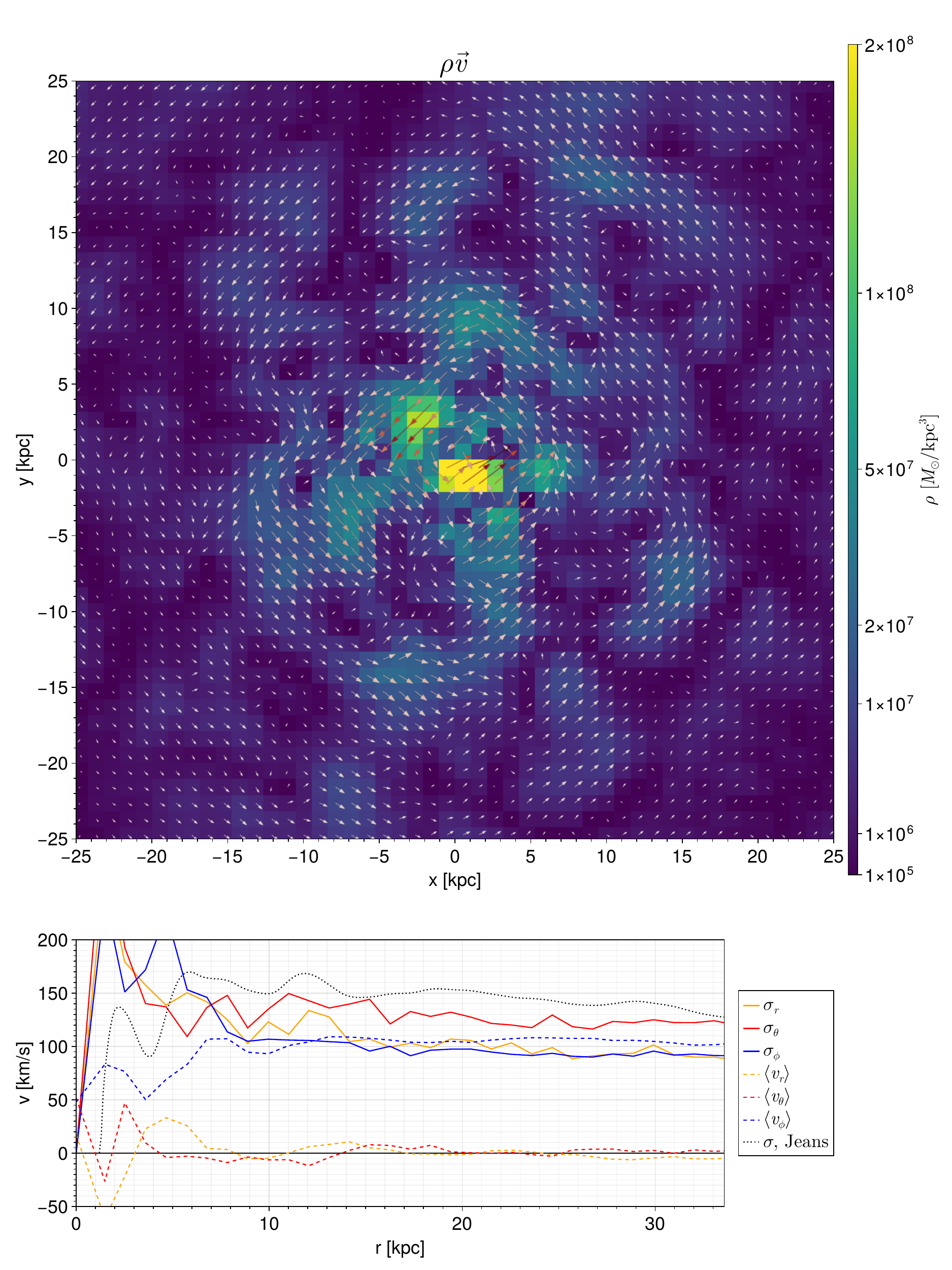}
    \caption{
        Velocity fields and dispersion profiles for the NGC~720 wave-CDM halo at the midplane at the final snapshot ($t=3$ Gyr).
        \textit{Top}: Momentum density vector field $(\rho v_x, \rho v_y)$ with density contours.
        The persistent azimuthal flow $\langle v_\phi \rangle \neq 0$ inherited from the initial conditions is visible in the tangential pattern.
        %%The lengths of arrows are cbrt-scaled.
        The arrow lengths are proportional to the cube root of the magnitude.
        %%% Figure A3, bottom:
        \textit{Bottom}: Radially binned statistics of spherical velocity decomposition: radial (orange), zenithal (red), and azimuthal (blue).
        Solid lines: shell-averaged mean velocities; dashed lines: one-dimensional velocity dispersions ($1\sigma$).
        Cyan curves: radial velocity dispersions predicted by the spherical Jeans equation (Eq.~\ref{eq:Jeans}).
        The agreement between the measured velocity dispersions and the Jeans estimate supports the interpretation that the halo has reached a virialized quasi-stationary state.
        The boson mass is $m_{22}=0.1$.
    }
    \label{fig:NGC720_vel_field}
\end{figure*}

The bottom panel of Fig.~\ref{fig:NGC720_vel_field}
compares the Jeans-predicted radial velocity dispersion with the simulated $\sigma_r(r)$ on the final snapshot.
The two profiles agree over the plotted radial range.
This is consistent with the halo having settled into a quasi-stationary state.

The top panel of Fig.~\ref{fig:NGC720_vel_field} shows the momentum density $(\rho v_x, \rho v_y)$ at the midplane.
The vector field shows a persistent azimuthal flow $\langle v_\phi \rangle \neq 0$,
which is inherited from the initial velocity-field setup.

\subsection{Temporal variation of the best-fit value of $a_0$}
 \label{sec:app_temporal-variation}
 
%% reviewer #2: time evolution of the RAR characteristic acceleration
As stated in the beginning of Sect.~\ref{sec:results},
a wave-CDM halo sustains persistent density fluctuations on the de~Broglie timescale (see \cite{hui2021wavedark}),
which cause the best-fit $a_0$ values to vary between snapshots even after virialization.
We therefore investigate the the temporal evolution of the
characteristic acceleration $a_0(t)$ for an NGC~720 simulation.
Certainly, this can also be viewed as a complementary diagnostic of both the approach to the NEQSS 
and the non-static character of the NEQSS.

\begin{figure*}[!ht]
    \centering
    \includegraphics[width=0.9\textwidth]{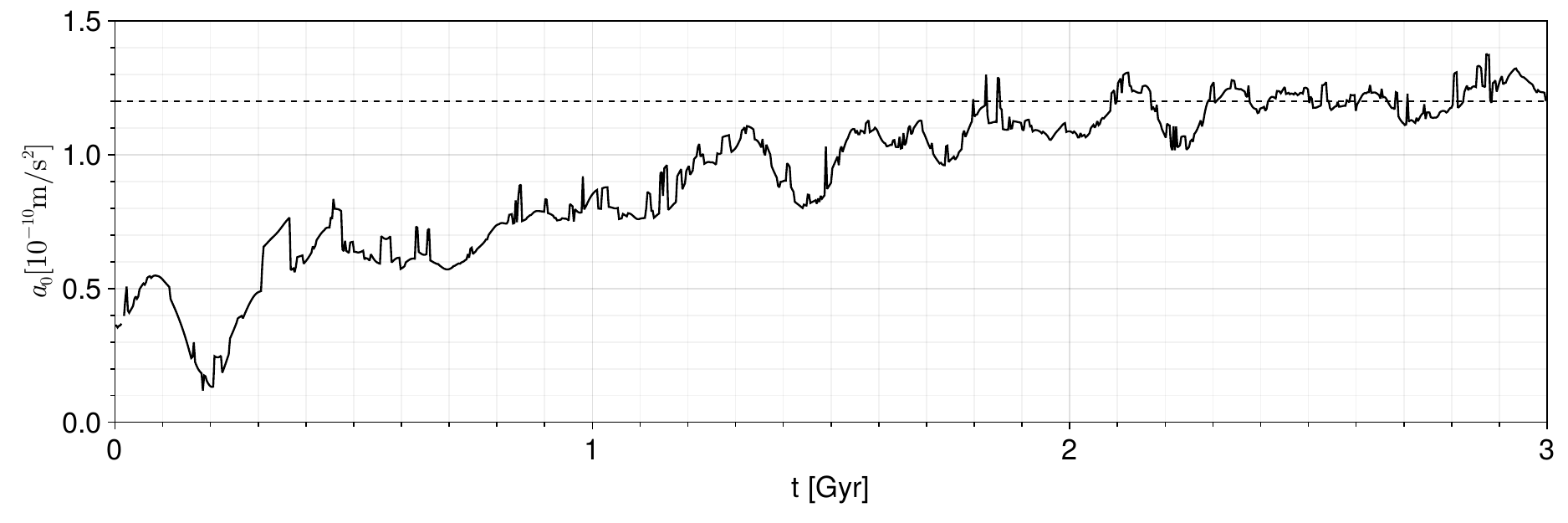}
    \caption{
        Temporal evolution of the best-fit RAR characteristic acceleration
        $a_0(t)$ for the NGC~720 wave-CDM simulation. %% with $m_{22}=0.1$.
        At each snapshot, $a_0$ is determined independently
        by fitting Eq.~\ref{eq:RAR}.
        The dashed horizontal line indicates the reference value
        $a_{0,\mathrm{ref}}=1.20\times10^{-10}\,\mathrm{m\,s^{-2}}$.
        Over the NEQSS interval $2\lesssim t/\mathrm{Gyr}\lesssim 3$,
        the temporal mean and standard deviation are
        $\langle a_0\rangle_t = 1.1963\times10^{-10}\,\mathrm{m\,s^{-2}}$
        and
        $\sigma_t(a_0)=0.0706\times10^{-10}\,\mathrm{m\,s^{-2}}$, respectively,
        corresponding to a relative dispersion of only
        $\sigma_t(a_0)/\langle a_0\rangle_t \sim 5.9\,\%$.
        The boson mass is $m_{22}=0.1$.
    }
    \label{fig:NGC720_a0_time}
\end{figure*}

%% At each snapshot, $a_0(t)$ is determined by fitting Eq.~\ref{eq:RAR}.
As shown in Fig.~\ref{fig:NGC720_a0_time}, 
during the initial relaxation phase ($t\lesssim 1\,\mathrm{Gyr}$),
$a_0(t)$ exhibits large temporal variations,
indicating that the system has not yet reached NEQSS.
After approximately $1.5\,\mathrm{Gyr}$,
the variations become substantially smaller,
and $a_0(t)$ remains close to the reference value $a_{0,\mathrm{ref}}
=1.20\times10^{-10}\,\mathrm{m\,s^{-2}}$,
shown by the dashed horizontal line.

Over the NEQSS interval $2\lesssim t/\mathrm{Gyr}\lesssim 3$,
the temporal mean and standard deviation are
$\langle a_0\rangle_t = 1.1963\times10^{-10}\,\mathrm{m\,s^{-2}}$
and
$\sigma_t(a_0)=0.0706\times10^{-10}\,\mathrm{m\,s^{-2}}$, respectively,
giving a relative dispersion of
$\sigma_t(a_0)/\langle a_0\rangle_t \simeq 5.9\,\%$ only.
This shows that, for the simulation considered here,
the time variation of $a_0$ during the NEQSS remains small.
The value of $a_0$ at the final snapshot ($t=3\,\mathrm{Gyr}$) is therefore representative of the NEQSS.
% The decreasing amplitude of the $a_0(t)$ fluctuations is consistent with the trend seen in
% the energy and mass-fraction diagnostics of
% Figs.~\ref{fig:app_energy}--\ref{fig:app_radii}.
%%} %\rev

\subsection{Mass profiles of the eight galaxies} \label{sec:app_mass_profile}

\begin{figure*}[!htbp]
    \centering
    \includegraphics[width=1.0\textwidth]{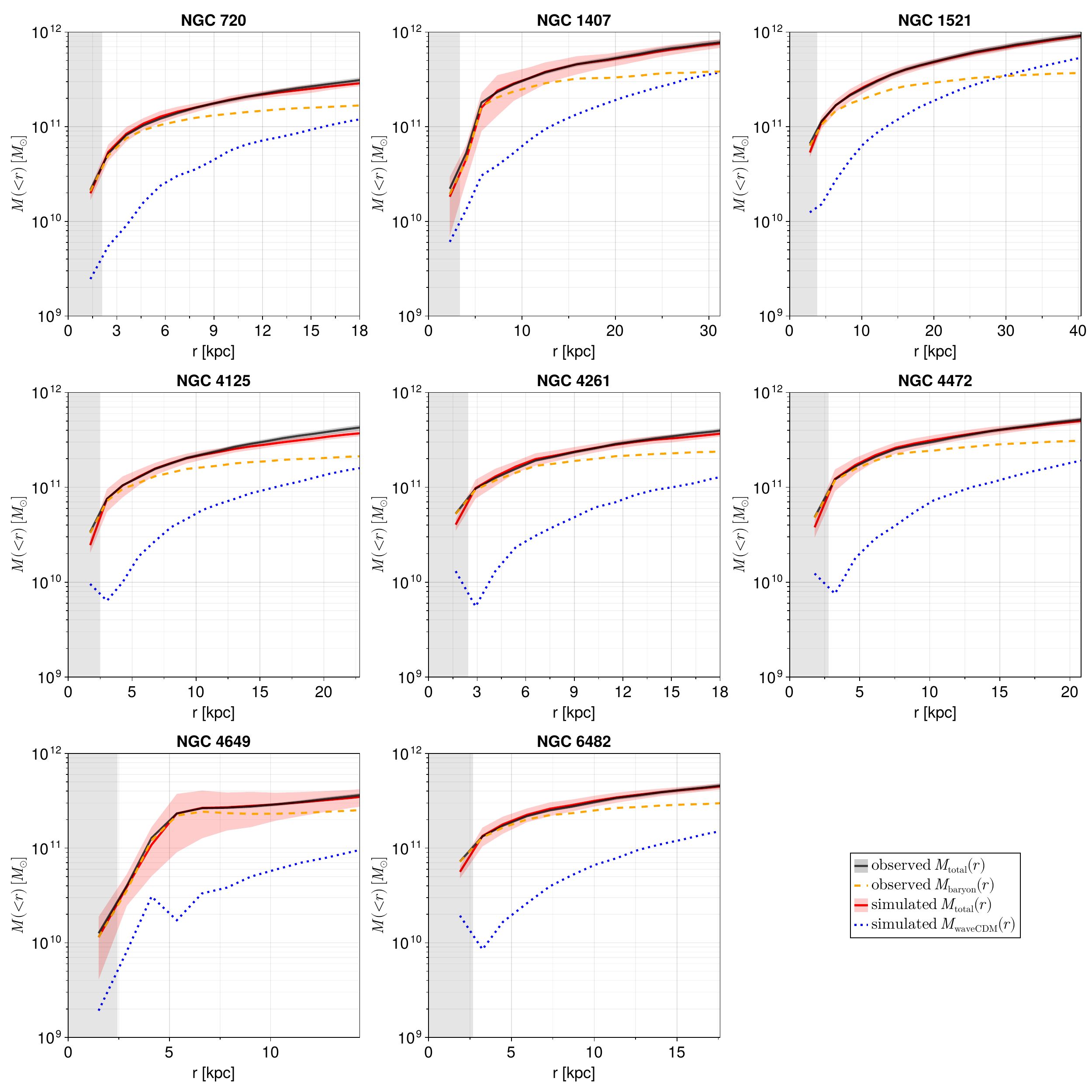}
    \caption{
        Enclosed-mass profiles, $M(<r)$, for the eight X-ray elliptical galaxies.
        For each galaxy, the observed total-mass profile \citep{lelli2017one} is compared with 
        that of the selected best-matching simulated realization at the final snapshot ($t=3\,\mathrm{Gyr}$; $m_{22}=0.1$).
        %%%%%
        Each panel extends out to the RAR outer cut at $4\,R_{\mathrm{eff}}$ (see Sect.~\ref{sec:simulations}).
        Black solid lines with shaded bands: observed total mass profiles $M_{\mathrm{total}}(r)$ from \cite{lelli2017one},
        with a $\pm 20\,\%$ band representing the systematic uncertainty from the stellar mass-to-light ratio $\Upsilon$.
        Orange dashed lines: observed baryonic component $M_{\mathrm{baryon}}(r)$ adopted from \cite{lelli2017one}.
        Red solid lines with shaded bands: simulated total (wave CDM and baryonic) mass profiles $M_{\mathrm{total}}(r)$,
        with the band giving the standard deviation of the azimuthal scatter within radial bins.
        Blue dotted lines: simulated wave-CDM halo component $M_{\mathrm{wave\,CDM}}(r)$.
        Gray shaded regions mark $r < \max(2\Delta x,\,0.1R_{\mathrm{eff}})$,
        excluded from the quantitative RAR analysis because of insufficient numerical resolution or central-baryon systematics.
        %For all eight galaxies, the simulated $M_{\mathrm{total}}(r)$ remain consistent with 
        %the observed profiles over the full plotted range.
    }
    \label{fig:app_mass_profile}
\end{figure*}

Fig.~\ref{fig:app_mass_profile} compares, for each of the eight X-ray elliptical galaxies, 
the observed radial mass profile with the simulated one at the baseline boson mass, $m_{22}=0.1$.
It provides the mass-profile counterpart to the RAR comparison shown in Fig.~\ref{fig:RAR_all_galaxies}.

%%% Appendix A5, July15:
\subsection{RAR fits for the combined samples of the simulated galaxies at different boson masses}
\label{sec:app_combined_rar}
%% moved to the Appendix, July14night:

Table~\ref{table:combined_RAR_all} lists the best-fit values of $a_0$ 
for the combined samples constructed, at each adopted boson mass, 
from the selected best-matching numerical realizations of all eight galaxies.
The fits were performed for $m_{22}=0.03$, $0.05$, $0.1$, and $0.2$.
%%%%  
The fitted values range from $1.06$ to $1.19\times10^{-10}\,\mathrm{m\,s^{-2}}$, 
remaining within approximately $11\%$ of the canonical value.

\begin{table}[H]
% \begin{table}[!htbp]
	\centering
	\footnotesize
    \caption{
	    Best-fit RAR parameters for the combined sample constructed from the selected best-matching numerical realizations
	    of the eight X-ray elliptical galaxies, for each adopted boson mass.
	    Column~(1) lists the boson mass;
	    column~(2) gives the best-fit value of $a_0$,
	    where the quoted uncertainty is the standard error returned by
	    the least-squares fit of Eq.~\ref{eq:RAR} to the combined radial data at the final snapshot,
	    which does not include the temporal variation of $a_0$ between snapshots,
	    the scatter among different initial-condition realizations,
	    or the systematic uncertainty ($\sim 20\,\%$) from the stellar mass-to-light ratio $\Upsilon$ \citep{lelli2017one};
        and column~(3) gives the number of data points included in every fitting.
   }
\smallskip
	\renewcommand{\arraystretch}{1.0}
	\begin{tabular}{ccc}
		\hline
		$m_{22}$ & $a_0$ ($10^{-10}\,\mathrm{m\,s^{-2}}$) & $N_{\mathrm{points}}$ \\
		\hline
		0.03 & $1.0641 \pm 0.0097$ & 4316 \\
		0.05 & $1.1702 \pm 0.0100$ & 3708 \\
		0.10 & $1.1903 \pm 0.0070$ & 6724 \\
		0.20 & $1.1891 \pm 0.0044$ & 14044 \\
		\hline
	\end{tabular}
	\label{table:combined_RAR_all}
\end{table}

\end{multicols}

\end{document}